# WIP: Assessing the Effectiveness of ChatGPT in Preparatory Testing Activities

Susmita Haldar
*School of Information Technology*
*Fanshawe College*
London, Canada
shaldar@fanshawec.ca

Mary Pierce
*Faculty of Business,*
*Info Technology and Pt Studies*
*Fanshawe College*
London, Canada
mpierce@fanshawec.ca

Luiz Fernando Capretz
*Department of Electrical and Computer Engineering*
*Western University*
London, Canada
lcapretz@uwo.ca

*Abstract*—This innovative practice WIP paper describes a research study that explores the integration of ChatGPT into the software testing curriculum and evaluates its effectiveness compared to human-generated testing artifacts. In a Capstone Project course, students were tasked with generating preparatory testing artifacts using ChatGPT prompts, which they had previously created manually. Their understanding and the effectiveness of the Artificial Intelligence generated artifacts were assessed through targeted questions. The results, drawn from this in-class assignment at a North American community college indicate that while ChatGPT can automate many testing preparation tasks, it cannot fully replace human expertise. However, students, already familiar with Information Technology at the postgraduate level, found the integration of ChatGPT into their workflow to be straightforward. The study suggests that AI can be gradually introduced into software testing education to keep pace with technological advancements.

*Index Terms*—Software Testing Education, ChatGPT, Black-box Testing, Product Testing, Higher Education

## I. INTRODUCTION

Software testing plays a critical role in delivering quality products to end users. The software that is not tested properly can exhibit unwanted critical bugs after moving to production which can cause a loss of goodwill, profit margin, and in turn shutdown of business [1], etc. To make this testing process efficient, researchers and practitioners are seeking innovative practices that can be applied in real-world applications.

Often organizations use third-party software or software as a service where source code is not accessible, leading the testers creating testing-related artifacts solely based on the specifications provided in the documentation. However, without examining the source code, software testers often face the challenge of ensuring proper requirements coverage of both functional and non-functional aspect.

Software testing educators are progressively adopting innovative tools and techniques to enhance the learning experience and equip students more effectively for real-world challenges. They need teaching students' effective ways of producing these testing entities. Traditional teaching methods often struggle to keep pace with the rapid advancements in software testing practices and technologies [2]. The software testing program curriculum should be updated to include the latest trends in modern technologies including Artificial Intelligence (AI), which can automate software testing processes, making them faster and more efficient. This integration will enable students to stay current with external developments in the testing field and contribute effectively to organizations after graduation [3].

Large Language Models (LLM) are a subset of AI and represent one of the latest inventions in deep learning and natural language processing (NLP) where the models are trained with vast amounts of data. ChatGPT is one of the LLM tools released by OpenAI [4] that allows users to use prompts with human-like language and receive the response on the provided prompts accordingly. Due to its simplicity, and human-like understanding of concepts, ChatGPT is showing promise in the various sectors of the Software Engineering domain including automation of software testing [5], [6].

This work-in-progress paper investigates the effectiveness of ChatGPT in preparatory testing activities without the need for having access to the source code of the application being tested. Also, this study identifies how post-graduate students in software testing program can apply their critical thinking skills in leveraging the benefits of LLM tools in generating testing preparatory activities while identifying the challenges of relying on these generated artifacts.

## II. BACKGROUND AND LITERATURE SURVEY

Several researchers compared the performance of different LLM tools including GPT-3.5 [7] and GPT-4 [8] and found ChatGPT-4 outperformed ChatGPT-3.5 [9]. ChatGPT-3.5 can ask for ChatGPT to answer follow-up questions, admit its mistakes, challenge incorrect premises, and reject inappropriate requests based on reinforcement learning from human feedback. In this study, ChatGPT-3.5 was used because of its free availability, and open-source tools are often preferred in educational settings due to being economical and creating room for innovation [10].

In the software testing paradigm, LLM tools have been experimented for unit test generation [11], test case generation from bug reports [12], GUI testing [13], code understanding [14], and program repair [15], etc. Wang et al. [16] conducted a review of the utilization of LLMs in software testing and identified test case preparation and program repair to be the most representative of software testing tasks.

TABLE I
LIST OF QUESTIONS USED IN THIS ASSIGNMENT

| Question Number | Question Description | Options provided |
| --- | --- | --- |
| 1 | In which areas did ChatGPT work the best compared to you working as a group in generating the questions manually? | a) Test Cases b) Test Scripts c) Use Cases d) RTM |
| 2 | Which aspect of ChatGPT-assisted testing did you find most beneficial? | a) Test case generation speed b) Test case coverage c) Test case accuracy d) Ease of use d) Other (please specify) |

Due to its popularity, and potential benefits in practical applications, the utilization of ChatGPT in the education domain is becoming inevitable. Mordan et al. [17] integrated LLM into higher education. Other researchers have incorporated ChatGPT into Software Engineering Education [18] and in Computing Education [19], etc.

Before the applications are ready for test execution, several activities are undertaken to enhance the efficiency of each step in the Software Testing Life Cycle (STLC) [20]. To keep track of the testing progress, and to measure the test coverage, the Requirements Traceability Matrix (RTM) links the requirements with the corresponding test cases. Madan et al. [21] showed the importance of RTM in testing web applications, and several other researchers worked on generating test cases from RTM [22]. There have been multiple studies on generating test cases from use case specification [23]–[25]. Considerable effort is required to generate test cases and test scripts [26]. Automating these steps would significantly reduce the total testing time otherwise required for a project.

## III. METHODOLOGY

This study has analyzed a subset of the student's responses to an assignment from the Capstone Project course of a post-graduate certificate in Software and Information Systems Testing Program from Fanshawe College, Canada. Students are required to complete the Capstone Project course in the second level of a two-semester program including a co-op term.

This Capstone Project course covers all aspects of testing a real-world web application. Students are expected to develop various testing-related artifacts over a 15-week duration semester. They need to develop use cases, RTM, test cases, and test scripts of a provided web-based TravelApplication as a group of 4-5 people. The students are evaluated based on their performance in these deliverables. In week 9, after manually completing all these activities, students are given an individual exercise to generate the same artifacts using ChatGPT. They then compare how manually generated test artifacts differ from those generated by ChatGPT. The students needed applying critical thinking, and problem-solving skills in reflecting their assessment of the generated artifacts. A total of 61 students submitted this assignment, and they were evaluated based on their findings. The students' responses that provided proper justifications and achieved a score of at least 90% in this assignment were utilized in this study. Table I shows the questions that were included in the exercise. Out of 11 questions, 2 were selected in this study as they were directly applicable to students' experience in generating testing preparatory artifacts compared to manual testing artifacts generation. 26 students got a score of above 90% in this assignment.

The students were given flexibility in using application information within the prompt, allowing them to learn how to tailor prompts to achieve the best outcome. This approach helped students understand that the non-deterministic nature of LLM tools may not always yield a specific outcome as easily as one might expect.

## IV. RESULTS

A few students initially prompted ChatGPT to generate Use Cases, RTM, Test Cases, and Test Scripts for the Travel Application without providing any additional context. The system returned information that was not very useful and generated a limited number of test cases and other artifacts. The students then adjusted their prompts applying their critical thinking skills to obtain the desired artifacts with more relevance.

### A. ChatGPT prompts by authors and responses received

This section shows a few examples of ChatGPT prompts used to generate preparatory testing artifacts.

Fig. 1 shows a snapshot of the prompt engineering applied by authors for providing the detailed specification in the prompt to get the required artifacts. In addition to what has been shown in this figure, the prompt included the description of the modules available in the existing application, the proposed application and details about the technology used in this TravelApplication. The Fig. 2 corresponds to one of the use cases generated by ChaGPT, which includes precondition basic, and alternate flow.

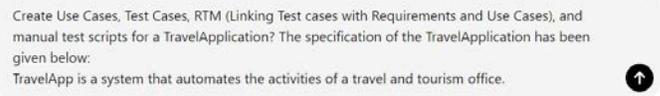

Fig. 1. Prompts for generating testing artifacts for a TravelApplication.

Fig. 3 demonstrates the generated RTM obtained from the given prompt. This RTM has been displayed in tabular format as instructed. Fig. 4 demonstrates how additional information had to be provided to obtain both functional and non-functional test cases. This prompt can be further enhanced to get a comprehensive set of non-functional test cases by specifically naming non-functional aspects of performance, load, stress, usability, learnability, and maintainability, etc.

Fig. 2. A ChatGPT-generated use case for login feature of the TravelApplication.

Fig. 3. Requirements Traceability Matrix generated by ChatGPT for the TravelApplication.

Fig. 4. Additional prompt for generating functional and nonfunctional test cases in a tabular format for the same TravelApplication

Fig. 5 shows functional and nonfunctional test cases generated from ChatGPT. However, because the prompt was too generic, specifying only general conditions instead of detailing which types of nonfunctional test cases to generate, the list of nonfunctional test cases was minimal.

*B. Students' findings from their assignment*

The student responses to the provided questions were analyzed afterward. Fig. 6 shows the students' responses on the question of which aspect of ChatGPT-assisted testing was most beneficial. Out of the 26 responses, fourteen students found that 'Test Case Generation Speed' was the most important aspect whereas four students considered the most benefit of ChatGPT-assisted testing is observed in the criteria of 'Ease of Use'. Three students referred to 'test case coverage' is the area where most benefit is observed compared to manual test artifacts generation. Two students selected the 'Test case generation speed and ease of use', two students picked 'Test Generation Speed and Test Case Coverage' and one student selected the 'Test Case Accuracy and Test Case' option as the best outcome of utilizing ChatGPT based test artifacts generation. Only one student selected 'Test Case Accuracy' as the most advantageous option. It was found that several students noted that the generated test cases were not always accurate, requiring them to tailor the prompts by providing additional context to obtain updated results.

Fig. 5. List of Functional and Non-Functional Test Cases

Fig. 7 shows which artifact among RTM, Use Cases, Test Cases, and Test Scripts was most effective according to the students and presents the results of this evaluation. Most students found use cases to be the most helpful artifact because they included all the pre-conditions, basic flows, and alternate flows. These alternate flows were sometimes missed in manual use case generation. RTM was the next preferred artifact as the ChatGPT generated responses demonstrated that each requirement was successfully linked to a specific test case. The students noted that some of the groups had inadvertently omitted non-functional test cases from their manually created testing artifacts, whereas these test cases were included in the ChatGPT-generated test cases. However, not all the ChatGPT-generated test cases apply to this application, which revealed ambiguities in the provided test specifications. Only three students found that the generated test scripts were helpful compared to others. As the application behaviors were not clearly defined in the specifications, the generated step-by-step instructions did not always match the actual requirements. Only those students who effectively tailored their requirements by providing more details in the ChatGPT prompt were able to achieve good results with their test script generation. Finally, a few students combined the effectiveness of use cases, RTM, and test scripts, by grouping them as either "use cases and RTM" or "test scripts and RTM". Two students found that the manually generated scripts were more effective than ChatGPT-generated artifacts shown as 'None' in this bar diagram.

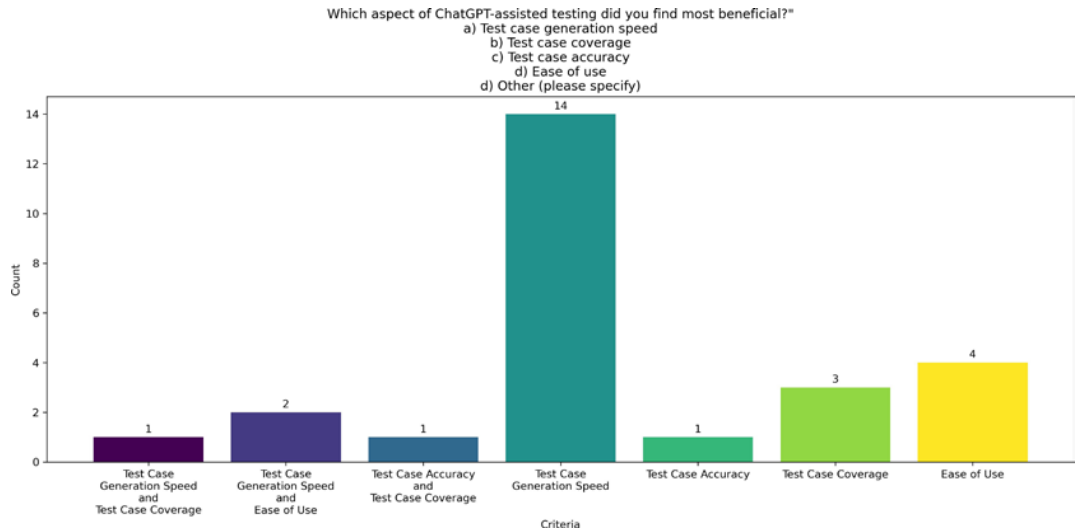

Fig. 6. Effectiveness of Testing Artifacts Generation

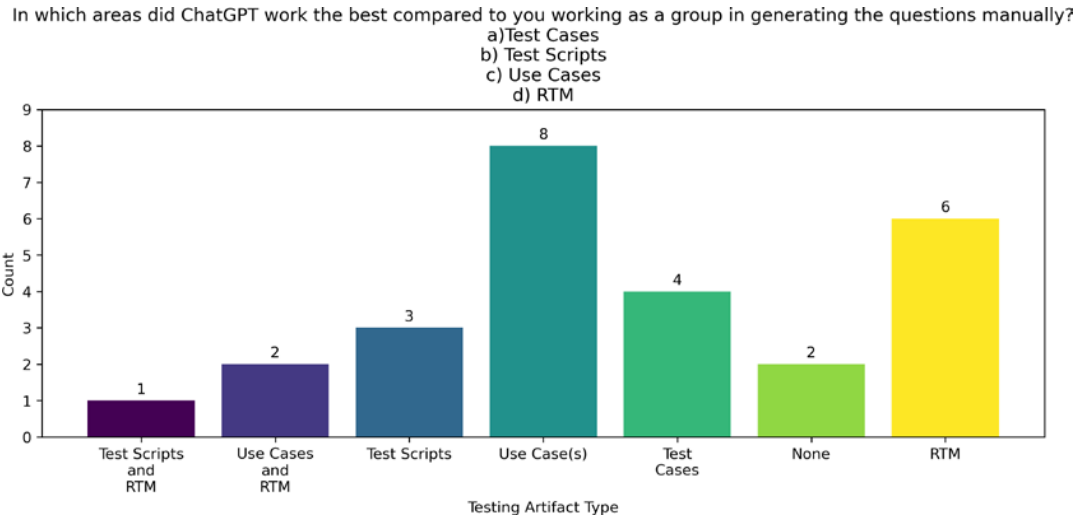

Fig. 7. Effectiveness of Testing Artifacts Generation

## V. DISCUSSION AND LIMITATIONS

From students' responses, it was observed that although ChatGPT automatically created RTM, use cases, test cases, and test scripts based on the given prompts, not necessarily all the generated information was correct or feasible to implement. ChatGPT has limitations such as the 'Software Under Test' cannot be directly accessed through a web browser. Without providing sufficient context in the prompt, the ChatGPT cannot generate an ample number of feasible test artifacts. To successfully generate test artifacts, providing a clear chain of instructions is crucial. In addition, when teaching with proprietary applications, there can be problems with sensitive information being shared as ChatGPT can use this for training. As a result, caution should be taken so that sensitive information is not shared unwillingly. The educators can show students different ways of prompt engineering to lead them in the right direction before handling this exercise. Since this set of twenty-six students was already in Information Technology domain in a post-graduate setting, it did not take time for them to familiarize themselves with test artifact generation using ChatGPT. Many of them have already utilized ChatGPT, but this exercise also made them aware that they cannot simply rely on ChatGPT without proper validation due to the risk of generating inaccurate information. Having prior knowledge with manual test artifacts generation can assist them with having the knowledge of creating effective prompts.

## VI. Conclusion and Future Direction

In this study, we explored how student assignments can be innovatively designed to compare their manually crafted testing artifacts with those generated by ChatGPT, providing a practical introduction to AI concepts within the testing domain. When integrating ChatGPT into the curriculum, guiding students in avoiding the inclusion of sensitive information in their prompts and to encourage them to validate the generated content before accepting the outcome is important. This innovative approach gradually introduces students to the complexities and challenges of using ChatGPT in real-world scenarios.

For future research, further analysis of student responses could yield valuable insights, and experimenting with additional open-source tools could offer a deeper understanding of ChatGPT's effectiveness in testing education. This continued investigation can help identify the potential and limitations of LLM tools in software testing studies.

## VII. Acknowledgement

We thank Dr. Dev Sainani, the associate dean of the School of Information Technology, Fanshawe College for his support in this research work.

## References

[1] I. S. T. Q. Board, *Certified Tester Foundation Level v4.0*, 2023. [Online]. Available: https://www.istqb.org/certifications/certified-tester-foundation-level

[2] O. A. Lazzarini Lemos, F. Fagundes Silveira, F. Cutigi Ferrari, and A. Garcia, "The impact of software testing education on code reliability: An empirical assessment," *Journal of Systems and Software*, vol. 137, pp. 497–511, 2018. [Online]. Available: https://www.sciencedirect.com/science/article/pii/S0164121217300419

[3] A. Rauf and M. N. Alanazi, "Using artificial intelligence to automatically test gui," in *2014 9th International Conference on Computer Science & Education*, 2014, pp. 3–5.

[4] OpenAI, "GPT-3 [OpenAI's ChatGPT]," https://openai.com/gpt-3/, 2020.

[5] J. Wang, Y. Huang, C. Chen, Z. Liu, S. Wang, and Q. Wang, "Software testing with large language models: Survey, landscape, and vision," *IEEE Transactions on Software Engineering*, vol. 50, no. 4, pp. 911–936, 2024.

[6] V. Bayrı and E. Demirel, "Ai-powered software testing: The impact of large language models on testing methodologies," in *2023 4th International Informatics and Software Engineering Conference (IISEC)*, 2023, pp. 1–4.

[7] T. Brown, B. Mann, N. Ryder, M. Subbiah, J. D. Kaplan, P. Dhariwal, A. Neelakantan, P. Shyam, G. Sastry, A. Askell *et al.*, "Language models are few-shot learners," *Advances in neural information processing systems*, vol. 33, pp. 1877–1901, 2020.

[8] OpenAI, "Introducing chatgpt-4," OpenAI Blog, March 2023, available online: https://openai.com/index/gpt-4/.

[9] J. López Espejel, E. H. Ettifouri, M. S. Yahaya Alassan, E. M. Chouham, and W. Dahhane, "Gpt-3.5, gpt-4, or bard? evaluating llms reasoning ability in zero-shot setting and performance boosting through prompts," *Natural Language Processing Journal*, vol. 5, p. 100032, 2023. [Online]. Available: https://www.sciencedirect.com/science/article/pii/S2949719123000298

[10] M. Dorodchi and N. Dehbozorgi, "Utilizing open source software in teaching practice-based software engineering courses," in *2016 IEEE Frontiers in Education Conference (FIE)*, 2016, pp. 1–5.

[11] M. Schäfer, S. Nadi, A. Eghbali, and F. Tip, "An empirical evaluation of using large language models for automated unit test generation," *IEEE Transactions on Software Engineering*, vol. 50, no. 1, pp. 85–105, 2024.

[12] S. Kang, J. Yoon, and S. Yoo, "Large language models are few-shot testers: Exploring llm-based general bug reproduction," in *Proceedings of the 45th International Conference on Software Engineering*, ser. ICSE '23. IEEE Press, 2023, p. 2312–2323. [Online]. Available: https://doi.org/10.1109/ICSE48619.2023.00194

[13] Z. Liu, C. Chen, J. Wang, M. Chen, B. Wu, X. Che, D. Wang, and Q. Wang, "Make llm a testing expert: Bringing human-like interaction to mobile gui testing via functionality-aware decisions," in *Proceedings of the IEEE/ACM 46th International Conference on Software Engineering*, ser. ICSE '24. New York, NY, USA: Association for Computing Machinery, 2024. [Online]. Available: https://doi.org/10.1145/3597503.3639180

[14] D. Nam, A. Macvean, V. Hellendoorn, B. Vasilescu, and B. Myers, "Using an llm to help with code understanding," in *Proceedings of the IEEE/ACM 46th International Conference on Software Engineering*, ser. ICSE '24. New York, NY, USA: Association for Computing Machinery, 2024. [Online]. Available: https://doi.org/10.1145/3597503.3639187

[15] M. Jin, S. Shahriar, M. Tufano, X. Shi, S. Lu, N. Sundaresan, and A. Svyatkovskiy, "Inferfix: End-to-end program repair with llms," in *Proceedings of the 31st ACM Joint European Software Engineering Conference and Symposium on the Foundations of Software Engineering*, ser. ESEC/FSE 2023. New York, NY, USA: Association for Computing Machinery, 2023, p. 1646–1656. [Online]. Available: https://doi.org/10.1145/3611643.3613892

[16] J. Wang, Y. Huang, C. Chen, Z. Liu, S. Wang, and Q. Wang, "Software Testing With Large Language Models: Survey, Landscape, and Vision," *IEEE Transactions on Software Engineering*, vol. 50, no. 4, pp. 911–936, Apr. 2024, conference Name: IEEE Transactions on Software Engineering. [Online]. Available: https://ieeexplore.ieee.org/abstract/document/10440574

[17] H. A. Modran, T. Chamunorwa, D. Ursuțiu, and C. Samoilă, "Integrating artificial intelligence and chatgpt into higher engineering education," in *Towards a Hybrid, Flexible and Socially Engaged Higher Education*, M. E. Auer, U. R. Cukierman, E. Vendrell Vidal, and E. Tovar Caro, Eds. Cham: Springer Nature Switzerland, 2024, pp. 499–510.

[18] V. D. Kirova, C. S. Ku, J. R. Laracy, and T. J. Marlowe, "Software engineering education must adapt and evolve for an llm environment," in *Proceedings of the 55th ACM Technical Symposium on Computer Science Education V. 1*, ser. SIGCSE 2024. New York, NY, USA: Association for Computing Machinery, 2024, p. 666–672. [Online]. Available: https://doi.org/10.1145/3626252.3630927

[19] J. Prather, P. Denny, J. Leinonen, B. A. Becker, I. Albluwi, M. Craig, H. Keuning, N. Kiesler, T. Kohn, A. Luxton-Reilly, S. MacNeil, A. Petersen, R. Pettit, B. N. Reeves, and J. Savelka, "The robots are here: Navigating the generative ai revolution in computing education," in *Proceedings of the 2023 Working Group Reports on Innovation and Technology in Computer Science Education*, ser. ITiCSE-WGR '23. New York, NY, USA: Association for Computing Machinery, 2023, p. 108–159. [Online]. Available: https://doi.org/10.1145/3623762.3633499

[20] F. Elberzhager, A. Rosbach, J. Münch, and R. Eschbach, "Reducing test effort: A systematic mapping study on existing approaches," *Information and Software Technology*, vol. 54, no. 10, pp. 1092–1106, 2012. [Online]. Available: https://www.sciencedirect.com/science/article/pii/S0950584912000894

[21] M. Madan, M. Dave, and A. Tandon, "Importance of rtm for testing a web-based project," in *2018 7th International Conference on Reliability, Infocom Technologies and Optimization (Trends and Future Directions) (ICRITO)*, 2018, pp. 816–818.

[22] B. Athira and P. Samuel, "Traceability matrix for regression testing in distributed software development," in *Advances in Computing and Communications*, A. Abraham, J. Lloret Mauri, J. F. Buford, J. Suzuki, and S. M. Thampi, Eds. Berlin, Heidelberg: Springer Berlin Heidelberg, 2011, pp. 80–87.

[23] C. Wang, F. Pastore, A. Goknil, and L. C. Briand, "Automatic generation of acceptance test cases from use case specifications: An nlp-based approach," *IEEE Transactions on Software Engineering*, vol. 48, no. 2, pp. 585–616, 2022.

[24] M. Jiang and Z. Ding, "Automation of test case generation from textual use cases," in *The 4th International Conference on Interaction Sciences*, 2011, pp. 102–107.

[25] C. T. M. Hue, D. D. Hanh, and N. N. Binh, "A transformation-based method for test case automatic generation from use cases," in *2018 10th International Conference on Knowledge and Systems Engineering (KSE)*, 2018, pp. 252–257.

[26] E. Dustin, T. Garrett, and B. Gauf, *Implementing automated software testing: How to save time and lower costs while raising quality*. Pearson Education, 2009.